\documentclass[%
 reprint,
superscriptaddress,
 amsmath,amssymb,
 aps,
prb,
floatfix]{revtex4-2}

\usepackage{graphicx}
\usepackage{dcolumn}
\usepackage{bm}

\begin{document}

\preprint{APS/123-QED}

\title{Exciton-Enhanced Superconductivity in Monolayer Films of Aluminum}

\author{Junhui Cao}
\affiliation{%
 Department of Physics, Zhejiang University, Hangzhou 310024, Zhejiang, China
}
\affiliation{%
 School of Science, Westlake University, Hangzhou 310024, Zhejiang, China
}%
\author{Alexey Kavokin}
\affiliation{%
 School of Science, Westlake University, Hangzhou 310024, Zhejiang, China
}%
\affiliation{%
 Abrikosov Center for Theoretical Physics, Moscow Center for Advanced Studies, Kulakova str. 20, Moscow, Russia
}%
\affiliation{%
 Russian Quantum Center, 30-1, Bolshoy boulevard, Skolkovo, Moscow Region, Russia.
}%


\date{\today}

\begin{abstract}
The BCS theory has achieved widespread success in describing conventional superconductivity. However, when the length scale reaches the atomic limit, the reduced dimensionality may lead to the quantum breakdown resulting in unpredictable superconducting behaviors. It has been experimentally evidenced that the critical temperature is strongly enhanced in the monolayer films of FeSe/STO and epitaxial aluminum. Here, we propose the exciton mechanism of superconductivity as a possible reason for the enhanced superconductivity in hybrid superconductor-semiconductor structures. The exciton-induced Cooper pairing may lead to the larger energy gaps and higher critical temperatures as compared to those caused by the phonon induced superconductivity. A detailed comparison of the theory and experimental results reveals the possibility of exciton-induced superconductivity in thin films of aluminum near the monolayer limit.
\end{abstract}

\maketitle


\section{Introduction}
The pursuit of higher-temperature superconductivity, particularly within the framework of Bardeen-Cooper-Schrieffer (BCS) theory\cite{PhysRev.108.1175}, has been a longstanding goal in condensed matter physics. Achieving this objective typically involves strategies such as the increase of a characteristic energy of a mediator of Cooper pairing in order to overcome the thermal fluctuation, and the enhancement of the coupling strength between mediators and electrons\cite{PhysRevLett.120.107001,RevModPhys.62.1027}. There exists a considerable interest in exploring systems out of thermal equilibrium, where alternative excitations in crystal, such as excitons, could potentially mediate electron pairing\cite{PhysRevLett.104.106402,PhysRevB.7.1020,CHEROTCHENKO2016170,CAO2023107293}. Although direct experimental evidence of exciton-mediated superconductivity remains elusive, discoveries of light-induced superconductivity have invigorated research in this area\cite{PhysRevLett.110.267003,science.1197294,PhysRevB.107.214506,science.1256783}. In these experiments, light generates crystal excitations akin to excitons, which facilitate electron pairing. In recent years, several research groups have considered the possibility of superconductivity mediated by a Bose-Einstein condensate of excitons\cite{PhysRevLett.104.106402} or exciton-polaritons\cite{PhysRevB.108.024513,PhysRevB.93.054510}. Building on the substantial progress in the experimental realization of bosonic condensates of exciton-polaritons at elevated temperatures\cite{sciadv.aau0244,PhysRevLett.98.126405,PhysRevLett.129.057402} these studies suggested the exciton-mediated superconductivity as a tool to achieve very high critical temperatures.

As superconductors approach atomic-scale dimensions, their properties can deviate significantly from predictions, leading to complex and unexpected behaviors\cite{science.aba5511,PhysRevLett.126.026802,accountsmr.4c00017}. In particular,  changes in dimensionality of the system and electron coupling to the surrounding environment can profoundly affect the electron-election pairing strength. While some materials exhibit a decrease in the critical temperature of superconductivity ($T_c$) as they approach the monolayer limit\cite{science.1170775}, others, such as FeSe on SrTiO$_3$, show a remarkable enhancement of $T_c$ in this regime\cite{027404-2025-0023,xu2021spectroscopic,song2021high}. Similarly, novel superconducting states can emerge at the interfaces of insulating materials, as demonstrated by the LaAlO$_3$/STO system\cite{science.1146006}. aluminum (Al), a classic type I BCS superconductor, displays unexpected modifications to its superconducting behavior when reduced to ultrathin films. It was reported in 2023 that the superconducting properties of epitaxial Al films grown on Si(111) substrates are greatly enchanced, as they approach the monolayer limit\cite{sciadv.adf5500}. Previous studies on Al films have reported widely varying $T_c$ values due to the differences in the preparation process of thin films, as well as experimental details such as the oxidation and granularity\cite{PhysRevLett.14.949,PhysRev.168.444}. One can conclude that while it is well known that $T_c$ can be increased in Al by reducing its thickness, the enhancement mechanisms remain unclear, particularly for films approaching the monolayer limit.

Here, we consider theoretically a two-dimensional hybrid semiconductor-superconductor structure, where electron-electron pairing in a superconductor can be mediated by excitons located in a semiconductor layer. In such hybrid structures, the combined effects of phonon coupling and exciton-mediated coupling, can be harnessed to achieve a substantial increase in the critical temperature of the superconductor. If the thickness of the aluminum film is reduced to monolayer limit, the role that the interface excitons play in mediating superconductivity becomes dominant. We present a straightforward model to illustrate how this mechanism could lead to larger gap and higher $T_c$ values.

\section{Model and Theory}
\subsection{Model}
\begin{figure}
\includegraphics[width=7 cm]{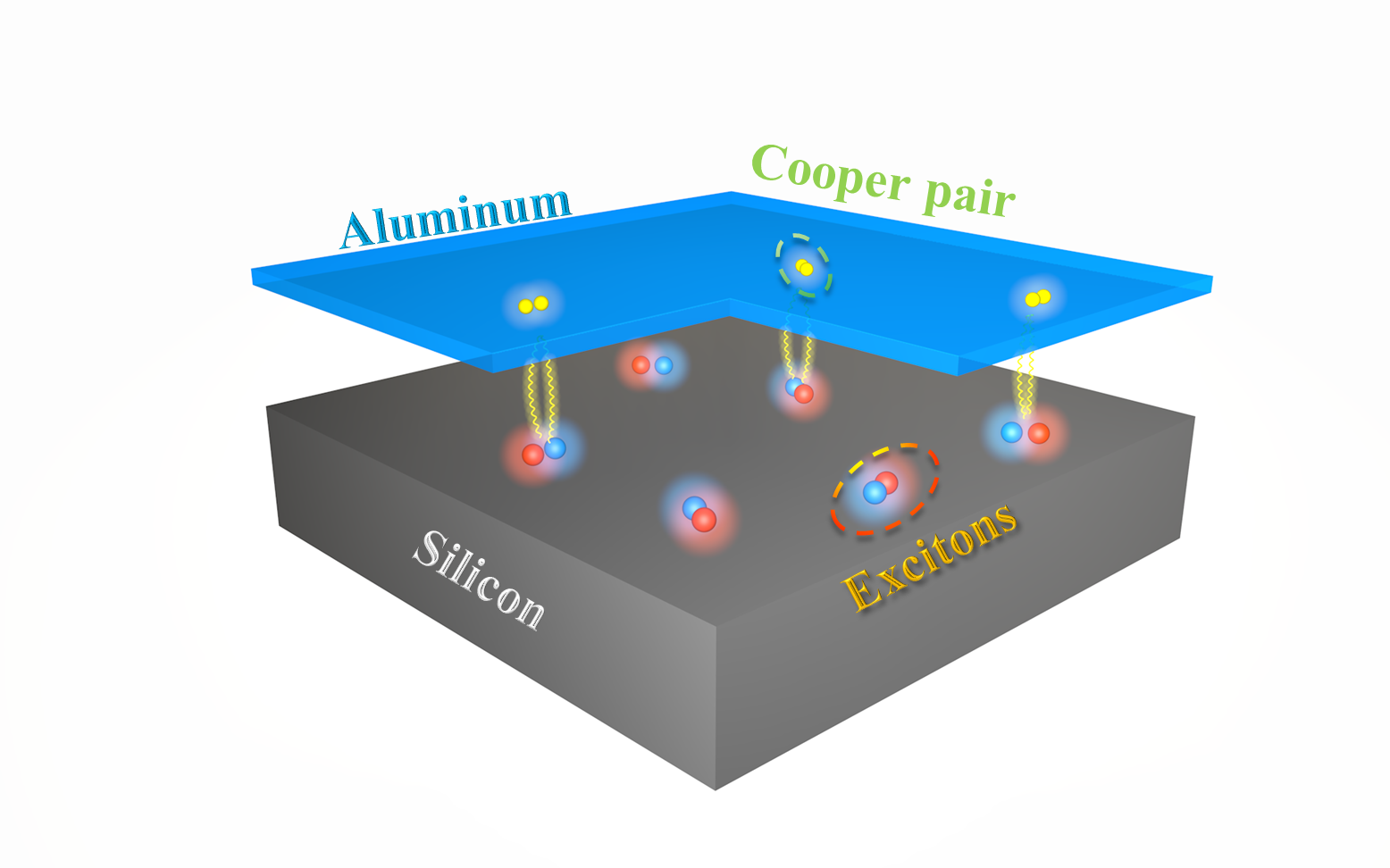}
\caption{Schematic of the model two-dimensional hybrid semiconductor-superconductor structure. The surface layer of silicon substrate sustains excitons (blue and red spheres represent the electron-hole pairs) due to the hole localization by the Schottky potential. Dipole-polarized excitons may mediate formation of electronic Cooper pairs (yellow spheres) in the adjacent aluminum layer, leading to the superconductivity. \label{fig1}}
\end{figure}

The model two-dimensional hybrid semiconductor-superconductor structures is illustrated in Figure \ref{fig1}. The bottom layer is a silicon substrate where excitons are formed in the vicinity of the surface due to the quantum confinement of holes in the Schottky potential. Delocalised electrons are bound to holes by the Coulomb attraction. As holes are much stronger localised at the surface than electrons, resulting excitons are dipole-polarised in the normal to the plane direction. Electrons in the ultrathin aluminum films interact with the polarised excitons in the silicon substrate, resulting in the effective attraction between free electrons, that form Cooper pairs. Due to the exponential decay of the screened Coulomb potential induced by dipole-polarized excitons along the c-axis (vertical axis), the formation of Cooper pairs mediated by excitons is only possible in very thin films of aluminum deposited on the silicon substrate. In aluminum films whose thickness exceeds 1-2 nm the electron-phonon interaction dominates over the excitonic mechanism. The interplay between phonon- and exciton-mediated Cooper pairing explains the increase of superconducting gap and $T_c$ with reduction of the thickness of the Al film grown on a silicon substrate that has been experimentally observed\cite{sciadv.adf5500}.

\subsection{Theory}
\begin{figure}
\includegraphics[width=7 cm]{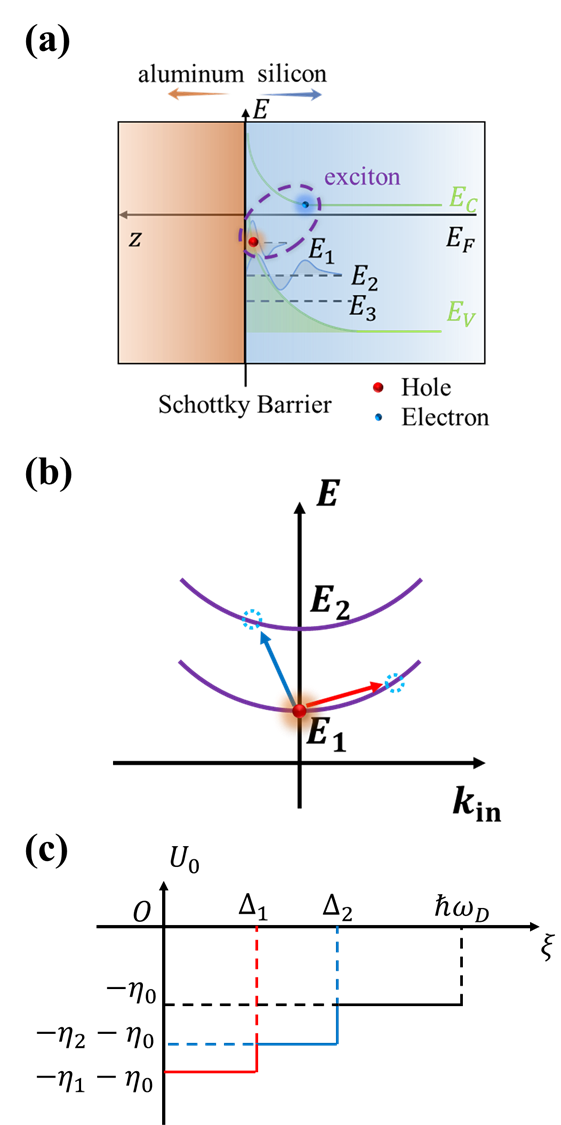}
\caption{a) Schematic of the confinement of excitons at the interface of silicon substrate and Al thin film. The green line indicates the potential created by the band bending effect. The horizontal dashed lines indicate the quantization of energy levels. Energies of conduction band, Fermi surface and valence band are denoted as $E_c, E_F$ and $E_v$; b) subband structure of the valence band, two hole scattering pathways are shown as the blue (inter-subband transition) and red (inner-subband transition) arrows; c) the three-step function showing the effective electron-electron interaction strength as a function of energy. \label{fig2}}
\end{figure}

Here we introduce the theory of exciton mediated formation of Cooper pairs in hybrid semiconductor-superconductor structures. Aluminum in such structure is p-type metal that creates a Schottky potential for both electrons and holes when deposited on the top of the silicon substrate \cite{Dingle1978665,Mahajan2021}. For holes, this potential is attractive, while for electrons it is repulsive. At sufficiently low temperatures, the electron-hole Coulomb coupling leads to the formation of spatially indirect, dipole-polarized excitons\cite{revuelta2023exciton} (Figure \ref{fig2}a). It is convenient to approximate the Schottky potential for holes by a triangular quantum well. The Hamiltonian for holes near the interface can be written as $H(z)=T+V(z)=-\frac{\hbar^2\nabla^2}{2m_h}+Fz$, where $m_h$ represents the mass of a hole and $F=0.5\ \text{meV}\cdot \text{nm}^{-1}$ is the magnitude of the built-in electric field. It is well-known that the general solution of the Schrodinger equation with a linear potential is Airy function:
\begin{equation}
\begin{aligned}
\varphi_i&=\frac\pi{\sqrt{3}}\sqrt{\xi_i}[J_{\frac13}(\frac23\xi_i^{\frac32})+J_{-\frac13}(\frac23\xi_i^{\frac32})],\\
\xi_i&=\left(\frac{2m_e}{\hbar^2F^2}\right)^{\frac13}(Fz-E_i),
\end{aligned}
\end{equation}
where $J$ stands for Bessel function, while the subscript $i$ indicates the subband considered. The corresponding eigen-energies $E_i$ of the first three hole subbands are $E_1=16.87\ \text{meV}, E_2=29.50\ \text{meV}$ and $E_3=39.84\ \text{meV}$, which are the ground state energies of hole subbands at the Si-Al interface.

Excitons mediate the interaction between two free electrons in the Al films as illustrated in Figure \ref{fig2}b). The scattering of a free electron with an exciton kicks the corresponding hole out of its ground state ($k_{in}=0$) at the lowest subband of the Schottky well and brings it to an excited state ($k_{in}\neq0$) of the same subband, or to one of the higher energy subbands  ($E_2, E_3,...$). To make the model as simple as it can be without losing the grasp of essential physics, we consider only hole excitations within the first two subbands. In two-dimensional system, if $L$ is small enough to validate the 2D electron gas (2DEG) assumption, the complex momentum- and energy-dependent exciton-induced electron-electron interacting strength (Eq.\ref{eqveff}) can be simplified as a constant value, given by E. Sedov and P. Skopelitis\cite{Sedov2020,PhysRevLett.120.107001}:
\begin{equation}
\eta=\frac{L^2d^2e^4n^2AN}{g_0\varepsilon^2}\label{eqconst}
\end{equation}
where $L$ is the average exciton-electron spacing dependent on the geometry of the potential well at the Si-Al interface, $d=1\ \text{nm}$ is the exciton dipole length\cite{AKMicrocavity}, $n$ is the exciton density, $A$ is the normalization area, $N=\frac{m}{\pi\hbar^2}$ is the density of states of the free electrons at the Fermi surface (2D case), $g_0$ is the exciton-exciton interaction constant ($g_0/A\sim\text{1 meV}$), $\varepsilon=15$ is the dielectric constant of silicon\cite{FRANTA2017405}. In the calculation, we have used the typical parameters of excitons in silicon\cite{PhysRevLett.95.156401,revuelta2023exciton}, besides, we set $L=0.5$ nm to describe the average separation of excitons and electrons in the hybrid structure. The value of the electron-electron interacting constant mediated by the transitions of holes in the first subband is $\eta_1=0.16$. The probability of hole excitation from the first to the second subband can be calculated with use of the Fermi's golden rule:
\begin{equation}
P_{1\to2}=\int\lvert\langle\varphi_2|\varphi_1\rangle\rvert^2D(E)dE,
\end{equation}
where $\varphi_1$ and $\varphi_2$ are the hole wavefunctions in the corresponding subbands, $D(E)$ is the hole density of states. The ratio of intra- and inter-subband transition rates is $\frac{P_{1\to2}}{P_{1\to1}}=0.747$, indicating the effective electron-electron interacting constant arising from the interband transition is $\eta_2=0.747*\eta_1=0.12$, while the phonon-induced effective coupling constant is $\eta_0=0.4$. Summing up these contributions, a three-step effective attractive interaction between free electrons can be plotted, as shown in Figure \ref{fig2}c. The energy cutoffs are denoted as $\Delta_1=E_2-E_1$ and $\Delta_2=E_3-E_1$, indicating the upper boundaries of the energy exchange for intra-subband ($1\to1$) and inter-subband ($1\to2$) transitions, respectively. In the original BCS theory, the formation of Cooper pairs mediated by phonons may only occur below the Debye temperature, and it is assumed to have a coupling strength independent of energy. In contrast, in the present model, below the first cutoff $\Delta_1$, Cooper pairs are generated by both electron-exciton scattering ($\eta_1=0.16$) and electron-phonon interaction ($\eta_0=0.4$). Between the first and second cutoffs, free electrons in Al will experience both the exciton inter-subband scattering ($\eta_2=0.12$) and phonon scattering. Finally, between the second and third cut-offs, only the phonon contribution to pairing is taken into account. The superconducting critical temperature and gap energy can be calculated numerically by solving the resulting gap equation:
\begin{equation}
\Delta(\xi,T)=\int_0^{\hbar\omega_D} \frac{U_0(\xi-\xi^{'})\Delta(\xi^{'},T)\mathrm{tanh}(\frac{E}{2k_BT})}{2E}\mathrm{d}\xi^{'},\label{gapeq}
\end{equation}
where $\Delta$ is the superconducting gap energy, $T$ the temperature, $U_0$ the effective electron-electron interaction strength corresponding to the exciton and phonon scattering mechanisms, $\hbar\omega_D=k_BT_D=35\ \text{meV}$ is the Debye energy of aluminum, and $E=\sqrt{\xi^{^{\prime2}}+\Delta^2(\xi^{^{\prime}},T)}$. 

Now, let us present the details of the quantum model used to derive the equations above.
To describe the exciton-induced attractive interaction between free electrons, we employ the standard random phase approximation (RPA) analysis. 
The second quantization Hamiltonian describing the interaction of electrons and excitons in the hybrid superconductor-semiconductor system reads:

\begin{widetext}
\begin{equation}
    H=\sum_{\mathbf{k}}[E_{\mathrm{el}}(\mathbf{k})a_{\mathbf{k}}^{\dagger}a_{\mathbf{k}}+E_{\mathrm{ex}}(\mathbf{k})b_{\mathbf{k}}^{\dagger}b_{\mathbf{k}}]+\sum_{\mathbf{k}_{1},\mathbf{k}_{2},\mathbf{q}}[V_{C}(\mathbf{q})a_{\mathbf{k}_{1}+\mathbf{q}}^{\dagger}a_{\mathbf{k}_{2}-\mathbf{q}}^{\dagger}a_{\mathbf{k}_{2}}a_{\mathbf{k}_{1}}+V_{X}(\mathbf{q})a_{\mathbf{k}_{1}}^{\dagger}a_{\mathbf{k}_{1}+\mathbf{q}}b_{\mathbf{k}_{2}+\mathbf{q}}^{\dagger}b_{\mathbf{k}_{2}}]+gn_\text{ex}
\label{eqHamil}
\end{equation}
\end{widetext}

Here $a_\mathbf{k}$ and $b_\mathbf{k}$ are annihilation operators for an electron and an exciton with a momentum $\mathbf{k}$, respectively. The first two terms represent the dispersion of the in-plane two-dimensional electrons gas ($E_{\mathrm{el}}$) and excitons ($E_{\mathrm{ex}}$) respectively. The third and forth terms describe the repulsive electron-electron Coulomb interaction ($V_{C}$) and electron-exciton interaction ($V_{X}$), with the momentum transfer between the interacting two particles labeled as $\mathbf{q}$. In our model, the relevant interaction is a Coulomb interaction between a free electron and a dipole-polarized exciton. As a result of this interaction, a hole bound to an electron to form the exciton is scattered with a free electron residing in the superconductor. The last term in Eq.~\ref{eqHamil} describes the exciton-exciton interaction, with $g=1\ \text{meV}\mu\text{m}^{2}$ being the short-range mean field interaction constant and $n_\text{ex}$ the exciton density. The detailed expression for $V_{C,X}$ writes\cite{JNP}:
\begin{widetext}
\begin{subequations}
\begin{align}
    V_{C}(\mathbf{q})&=e^2/[2\epsilon A(|\mathbf{q}|+\kappa)],\\
    V_{X}(\mathbf{q})& =\frac{e^2}{2\epsilon A}\frac{e^{-|\mathbf{q}|L}}{|\mathbf{q}|}\left\{\frac{1}{[1+(\beta_e|\mathbf{q}|a_B/2)^2]^{3/2}}-\frac{1}{[1+(\beta_h|\mathbf{q}|a_B/2)^2]^{3/2}}\right\} \notag \\
    &+\frac{e^2d}{2\epsilon A}e^{-|\mathbf{q}|L}\left\{\frac{\beta_e}{[1+(\beta_e|\mathbf{q}|a_B/2)^2]^{3/2}}+\frac{\beta_h}{[1+(\beta_h|\mathbf{q}|a_B/2)^2]^{3/2}}\right\},
\end{align}
\label{eqHamCX}
\end{subequations}
\end{widetext}
\noindent
where $\kappa=m_ee^2/(2\pi\epsilon\hbar^2)$, with $\epsilon=7\epsilon_0$ being the dielectric constant of the screened Coulomb potential\cite{PhysRevLett.104.106402,PhysRevB.93.054510}, $\beta_{e,h}=m_{e,h}/(m_e+m_h)$ is the mass ratio of electron/hole in the exciton, here we take $\beta_{e}=0.1$ with $m_e=0.03 m_0$, $n_{\text{ex}}=10^{12}$ cm$^{-2}$ the exciton density, which can be estimated from the work functions of metal and semiconductor, and $d=1\ \text{nm}$ is the average separation of electron and hole along the structure axis that defines the exciton stationary dipole moment\cite{AKMicrocavity}. As we mentioned, excitons are localized and, at low temperatures, occupy mainly the ground state. Therefore, for exciton operators we have  $\mathbf{k}_2=0$, and $\langle b_{\mathbf{k}}^\dagger\rangle\approx\langle b_{\mathbf{k}}\rangle\approx\sqrt{n_{\mathrm{ex}}A}\delta_{\mathbf{k},\mathbf{0}}$, which is the population of excitons at the ground state. 
We emphasize that in the system under consideration, the exciton gas at the interface is formed without external pumping, as a result of band bending by the Schottky potential. Excitons are stable in the low temperature limit as shown in Ref.\cite{revuelta2023exciton}. By substituting the bosonic operator mean-field approximation $b_{\mathbf{k+q}}^\dagger b_\mathbf{k}\approx\langle b_{\mathbf{k+q}}^\dagger\rangle b_\mathbf{k}+b_{\mathbf{k+q}}^\dagger\langle b_\mathbf{k}\rangle$ into the four-operator expression in Eq.~\ref{eqHamil}, now the electron-exciton interaction term can be simplified as three-operator expression: $\sqrt{N}V_{X}(\mathbf{q})\Sigma_{\mathbf{k},\mathbf{q}}a_{\mathbf{k}}^{\dagger}a_{\mathbf{k}+\mathbf{q}}(b_{-\mathbf{q}}^{\dagger}+b_{\mathbf{q}})$, with $N=n_{\text{ex}}A$. This expression is similar to the familiar electron-phonon interaction term. Given that a Cooper pair is formed due to the subsequent forward and backward scattering of the exciton hole, the similarity between exciton and phonon mechanisms is apparent. Due to the bosonic nature of excitons, a propagator of an exciton with momentum $\mathbf{q}$ can be written as:
\begin{equation}
D(\mathbf{q},\omega)=\frac{2\omega_\mathbf{q}}{\omega^2-\omega_\mathbf{q}^2+i\delta},
\end{equation}
here, a quadratic dispersion of exciton kinetic energy $\omega_\mathbf{q}$ is proposed. The effective interaction mediated by excitons is:
\begin{equation}
V_{\mathrm{eff}}(\mathbf{q},\omega)=V_X(\mathbf{q})^2D(\mathbf{q},\omega).
\label{eqveff}
\end{equation}
In the low-frequency limit,$V_{\mathrm{eff}}\approx-V_X(\mathbf{q})^2/\omega_\mathbf{q}$, is negative, which explains the attractive interaction between electrons mediated by excitons in the low-energy excitation region. Once the exciton-mediated attraction overcomes the repulsion between electrons, Cooper pairs can be formed, and the system enters the superconducting state. The total effective interaction potential $V(\mathbf{q},\omega)$ can be obtained by summing up the series of bubble diagrams representing electron-electron screening. This involves both the direct Coulomb interaction and the exciton-mediated interaction, that is:
\begin{equation}
V(\mathbf{q},\omega)=\frac{V_{{\mathrm{C}}}(\mathbf{q})+V_{{\mathrm{eff}}}(\mathbf{q},\omega)}{1-\chi_0(\mathbf{q},\omega)\left[V_{{\mathrm{C}}}(\mathbf{q})+V_{{\mathrm{eff}}}(\mathbf{q},\omega)\right]},
\end{equation}
with
\begin{equation}
\chi_0(\mathbf{q},\omega)=\sum_{\mathbf{k}}\frac{f(\epsilon_{\mathbf{k}})-f(\epsilon_{\mathbf{k}+\mathbf{q}})}{\omega+\epsilon_{\mathbf{k}}-\epsilon_{\mathbf{k}+\mathbf{q}}+i\delta},
\end{equation}
being the polarization function of electrons, and $f$ being the Fermi-Dirac distribution for electrons. When $V(\mathbf{q},\omega)<0$, electrons in the thin films of metal experience an effective attraction mediated by excitons, leading to the formation of Cooper pairs and superconductivity. The mechanism of Cooper pairing with excitons is essentially the same as one with phonons, however, it strongly depends on the geometry of the hybrid structure, being highly efficient in the regime of strong proximity between superconducting and excitonic layers. The increase in the critical temperature of Al thin films with the reduction of dimensionality is a consequence of exciton-mediated electron-electron attractive interaction, while the superconductivity in a bulk system is dominated by the phonon contribution. The dimensionless effective electron-electron interaction potential $U_{\text{eff}}(\omega)$ can be written as:
\begin{equation}
U_{\text{eff}}(\omega)=\frac{A\mathcal{N}}{2\pi}\int_0^{2\pi}V(q,\omega)d\theta,
\label{eqeff}
\end{equation}
with $q=k_F\sqrt{2(1-\cos\theta)}$ and $\mathcal{N}=\frac{m_e}{\pi\hbar^2}$ the density of states of the 2DEG of aluminum thin film. The results of our calculations are summarized in Figure 3 and discussed below.

\begin{figure}
\includegraphics[width=6 cm]{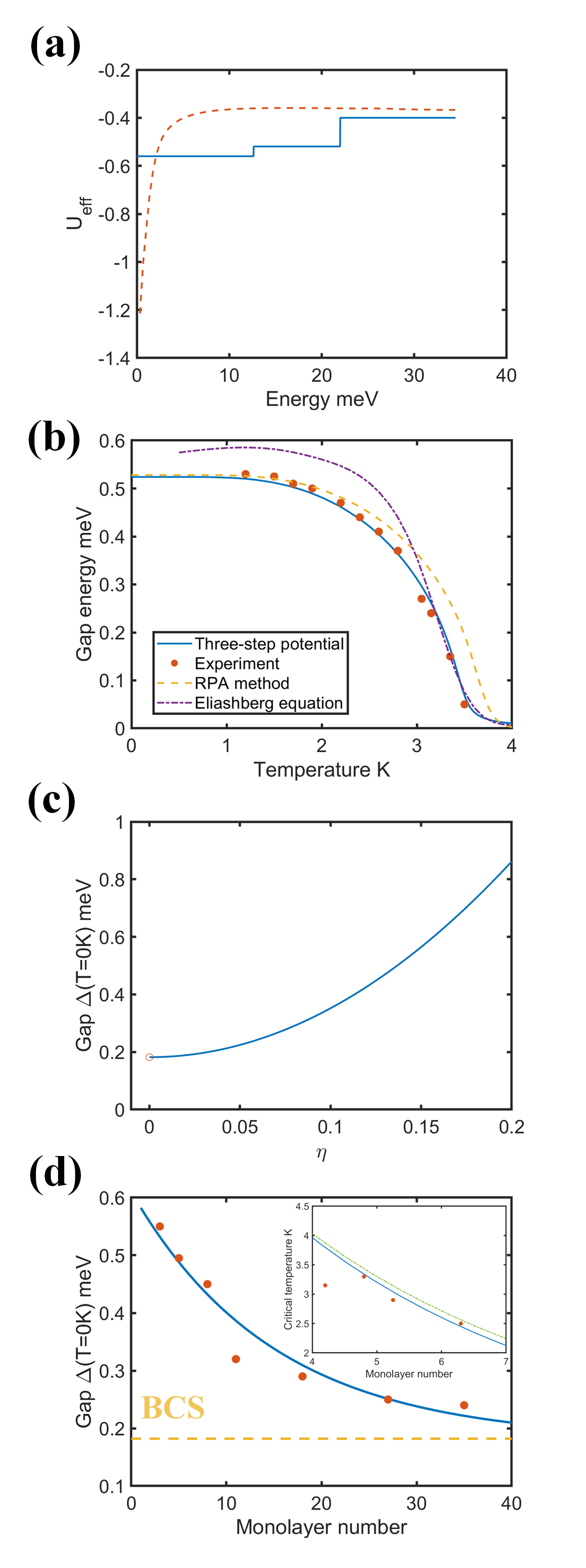}
\caption{a) The effective dimensionless interaction potentials from the RPA calculation (Eq.~\ref{eqeff}, dashed curve) and the simplified three-step potential model (solid curve). b) The temperature dependence of the superconducting energy gap. The solid blue curve is the result of numerical calculation from the three-step potential model, the dashed yellow curve is the result from the RPA calculation, the violet dash-dotted curve is obtained by solving the imaginary axis Eliashberg equation, while red circles are experimental data extracted from Ref.\cite{sciadv.adf5500}. c) Dependence of the superconducting energy gap on the interaction strength $\eta$. The red circle indicates the purely phonon contribution, with a critical temperature of 1.2 K. d) The dependence of gap energy on the thickness of Al layer (in monoatomic layers). Red dots are the experimental data adapted from \cite{sciadv.adf5500}; solid curve is the theoretical result; dashed line indicates the bulk phonon contribution. Inserted: critical temperature versus Al thin film layer number. Red dots are extracted from experiment data \cite{sciadv.adf5500}, blue solid curve is the result from RPA, and the green dash-dot curve is calculated from McMillan equation.
\label{fig3}}
\end{figure}

\section{Result and Discussion}
In Figure~\ref{fig3}a, we compare the simplified three-step potential model with the results of the exact solution of the gap equation based on the effecting electron-electron interaction potential introduced above from the RPA analysis, with a constant background contribution ($U=-0.4$) from the electron-phonon interaction. The negative value of $U_{\text{eff}}$ in the low-energy exchange regime can be ascribed to the attractive interaction mediated by the excitons, which leads to the formation of Cooper pairs in the aluminum thin films. With the increase of the energy exchange, the repulsive Coulomb interaction gradually becomes dominant due to the reduction of the exciton propagator $D$, resulting in an increased $U_{\text{eff}}$. In comparison, the three-step potential model grasps the increasing tendency of the increased electron-electron interaction with the energy exchange while largely reduces the calculation cost and 
can be easily understood.

Figure \ref{fig3}b shows the temperature dependence of the superconducting energy gap. Solid line is a result of the numerical solution of Eq.\ref{gapeq}, where an iteration method has been employed and the initial constant gap function has been used as a trial function. In this model, the superconducting gap closes at the critical temperature about 3.4 K. This is very high compared to the bulk aluminum $T_c$ (1.2 K). The exciton contribution to the Cooper pairing leads to the increase of $T_c$ by a factor of 2.8. This theoretical result is in the excellent agreement with the experimental data (red circles) reported in Ref.\cite{sciadv.adf5500} which supports the hypothesis of exciton-mediated superconductivity in the silicon-Al heterostructure. While BCS theory offers a foundational description of superconductivity through static electron-electron interactions, it neglects the dynamical retardation effects inherent in electron-boson coupling. To account for these effects, we employ both frequency-dependent RPA and imaginary-axis Eliashberg calculations. The dashed curve is obtained from the RPA calculation, which has an almost identical gap energy (0.51 meV) to the three-step potential model at zero temperature. However, at finite temperatures, it shows a slightly overestimated gap energy, which might be related to the underestimated $V_C$ due to the underestimated screening effect in low dimension\cite{2015Screening,10.1063/5.0132820} as well as the enhancement from the retardation effect due to the resonance near the exciton frequency. The imaginary-axis Eliashberg gap (violet dash-dotted curve, details in Appendix) incorporates retardation via the frequency-dependent coupling parameter $\lambda(i\omega_m)$. In the high temperature regime, the prediction from the imaginary axis Eliashberg equation is in accordance with the three-step potential model, reflecting the weak-coupling behavior, consistent with the small $\lambda_0\ll10$. While in the low temperature regime, on account of the dynamical interaction between electrons and phonons or excitons, the gap function predicted by the Eliashberg equation is slightly higher. Crucially, both RPA and Eliashberg results suggest that retardation effects are minor in this system. This validates the three-step potential model as an accurate and convenient description of exciton-mediated pairing at the Al-Si interface.

Further, we examined the dependence of the gap energy on the strength of electron-electron interaction mediated by excitons (Eq.\ref{eqconst}). It is shown in Figure \ref{fig3}c. As expected, the superconducting energy gap increases with the increase of the coupling strength. A robust superconducting state may be formed once the interaction strength is large enough. In practice, the interaction constant can be tuned by tuning the density of excitons in the silicon substrate with use of an external bias. The external bias would be added up to the Schottky potential altering the hole density and, consequently, the exciton density. In turn, the superconducting gap can be enhanced or reduced depending on the density of active excitons in the surface layer of the substrate. This paves the way towards a variety of applications such as superconducting transistors or qubits. 

The calculated dependence of the gap on the width of the aluminum film is shown in Figure \ref{fig3}d, which is obtained by averaging
the bulk contribution coming from the electron-phonon interaction and the interface contribution coming from both electron-phonon and electron-exciton interactions. Due to the finite interaction length ($L=0.5$ nm, one monolayer) between excitons in Si substrate and electrons in Al thin film, only Cooper pairs at the interface can be formed from the benefit of exciton-electron interaction. With an increase in the number of Al layers, the critical temperature approaches the limit of the bulk aluminum metal. This strong dependence of the gap on the aluminum film thickness can be considered as a smoking gun for the exciton mechanism of superconducting. Further in the inserted graph, Migdal-Eliashberg theory \cite{migdal1958interaction,eliashberg1960interactions} combined with the McMillan equation \cite{PhysRev.167.331} is used to study the electron-exciton interaction in the exciton mechanism. Details about the calculation can be found in the appendix. In the numerical calculation, the electron-exciton interaction constant $\lambda\in[0.745,0.851]$ and $\mu^*=0.108$ are within the applicable scope of the McMillan equation where $\lambda\leq10$ and $\mu^*\leq0.2$. The constant $\lambda$ for the electron-boson interaction is higher than that of the electron-phonon interaction ($\lambda^*=0.42$, data from Ref.~\cite{PhysRevB.54.16487}) in Al, which can explain the increase in the critical temperature in the Al thin film within the framework of the exciton mechanism. Due to the isotropic nature of the electron-exciton interaction $V_X$, the result derived from the Migdal-Eliashberg theory is almost identical to the result obtained from the RPA using Eq.~{\ref{eqeff}}. However, on account of the renormalization to the screened Coulomb interaction with the Morel-Anderson pseudo-potential, the $T_c$ estimated from the McMillan equation (green dash-dotted curve) is slightly larger than the calculation from RPA (blue solid curve).


\section{Conclusion}
In summary, we have considered the exciton mechanism as a possible cause of the enhanced superconductivity in thin films of aluminum near the atomic limit, where the excitons located in semiconductor substrate may serve as mediators of electron-electron coupling. The transition from a bulk superconductor to an atomically this layer of superconductor witnesses the reduction of the relative phonon contribution to the Cooper pairing, while the role of the exciton-mediated electron-electron interaction is flaring up. It is worth mentioning that the formation of Cooper pairs is considered here in the framework of conventional $s$-wave pairing. Our model opens a new way to explain the enhanced superconductivity in the systems of reduced dimensionality and, in particular, at the interfaces of semiconductor and superconductor heterostructures. While the critical temperature of superconductivity in such structures is strongly dependent on the density of excitons that can be controlled with applied bias or by means of optical excitation. The dependence of the critical temperature on the exciton concentration would be a fingerprint of the exciton-induced superconductivity. Given the increasing reliance on integrating superconductors into heterostructures for quantum technologies, such as superconducting spintronics, high-precision magnetometers, and superconducting qubits, we believe the discovery of exciton-induced superconductivity would further boost the development of superconductors.

\section{Acknowledgement}
We acknowledge the National Science Foundation of China for the support within the project W2431003 and Innovation Program for Quantum Science and Technology 2023ZD0300300. J. C. appreciates the kind support from Prof. Congjun Wu.

\appendix*

\section{Eliashberg spectral function and McMillan equation}
Migdal-Eliashberg theory is a microscopic framework which describes conventional superconductivity mediated by electron-phonon interactions. It generalizes BCS theory by incorporating strong coupling effects and the retarded nature of phonon-mediated pairing. In the exciton mechanism, the three-operator form of the electron-exciton interaction term allows one to analyze it in analogy with the electron-phonon interaction from the Migdal-Eliashberg theory. The spectral function $\alpha^2F$ is used to estimate the electron-exciton interaction strength, with its peaks corresponding to exciton modes that strongly mediate Cooper pairing, which is defined as:
\begin{equation}
\alpha^2F(\omega)=\frac{N}{N(0)}\sum_{k,k^{\prime}}\frac{V_X^2(q)}{\epsilon(q)}\delta(\omega-\omega_q)\delta(\varepsilon_k-\varepsilon_F)\delta(\varepsilon_{k^{\prime}}-\varepsilon_F),
\end{equation}
where $N(0)$ is the density of states in the 2DEG, $V_X(q)$ is the electron-exciton interaction (Eq.~\ref{eqHamCX}b), $\epsilon(q)=1-V_C(q)\chi_0(q)$ is the dielectric function derived from RPA. $q=k-k^\prime$ is the momentum transfer, $\omega_q$ is the dispersion of the free exciton, and the Dirac function indicates that scatterings away from the Fermi energy $\varepsilon_F$ are forbidden.

The linewidth $\gamma_q$ of the exciton can be read from the imaginary part of the polarization function $\chi_0$, which is
\begin{equation}
    \begin{aligned}
        \gamma_q&=\frac{2NV_X^2(q)}{\epsilon^2(q)}\mathrm{Im}[\chi_0(q,\omega_q)]\\
        &=\frac{2\pi NV_X^2(q)}{\epsilon^2(q)}\Sigma_k[f(\varepsilon_k)-f(\varepsilon_{k+q})]\delta(\omega_q-\varepsilon_k+\varepsilon_{k+q}).
    \end{aligned}
\end{equation}
The second line can be obtained with the identical equation $\mathrm{Im}[\frac{1}{\omega+i\delta-\varepsilon}]=-\pi\delta(\omega-\varepsilon)$ derived from residue theorem. By using the approximation $f(\varepsilon_k)-f(\varepsilon_{k+q})\approx\omega_q\delta(\varepsilon-\varepsilon_F)$, and assuming the interaction to be isotopic, finally we will arrive at the expression of $\gamma_q$ that
\begin{equation}
\begin{aligned}
            \gamma_q&\approx\frac{2\pi NV_X^2(q)}{\epsilon^2(q)}\Sigma_k\delta(\varepsilon_k-\varepsilon_F)\delta(\varepsilon_{k+q}-\varepsilon_F)\\
            &=\frac{\omega_qNV_X^2(q)N(0)}{\varepsilon_F\epsilon^2(q)}\frac{2k_F}{q\sqrt{1-(q/2k_F)^2}}.
\end{aligned}
\end{equation}
The spectral function can be expressed with $\gamma_q$, which is
\begin{equation}
\alpha^2F(\omega)=\frac{1}{2\pi N(0)\omega}\sum_q\gamma_q\delta(\omega-\omega_q).
\label{a2f}
\end{equation}
The dimensionless parameter $\lambda$ quantifies the average strength of electron-exciton interactions in this system, which is defined as
\begin{equation}
\lambda=2\int\frac{d\omega\alpha^2F(\omega)}{\omega}=\frac{1}{\pi N(0)}\sum_q\frac{\gamma_q}{\omega_q^2}.
\end{equation}
Since the scattering occurs only around the Fermi surface, the summation over momentum transfer $q$ can be treated as the integral from $q=0$ to $q=2k_F$. Therefore, the analytical expression of the electron-exciton interaction constant $\lambda$ is then
\begin{equation}
\lambda=\frac{2N(0)}{\pi k_F}\int_0^{2k_F}dq\frac{NV_X^2(q)}{\omega_q\epsilon^2(q)}\left[1-\left(\frac{q}{2k_F}\right)^2\right]^{-1/2}.
\label{lambda}
\end{equation}

The critical superconducting temperature $T_c$ in Migdal-Eliashberg theory can be obtained in an elegant way by the McMillan equation with Allen-Dynes correction, which is
\begin{equation}{
\begin{aligned}
k_{B}T_{c} & =\frac{f_1f_2\omega_{\log}}{1.2}\exp\left[-\frac{1.04(1+\lambda)}{\lambda(1-0.62\mu^*)-\mu^*}\right], \\
f_{1} & =[1+(\lambda/\Lambda_1)^{3/2}]^{1/3},\\ f_2&=1+\frac{({\omega}_2/\omega_{\log}-1)\lambda^2}{\lambda^2+\Lambda_2^2}, \\
\Lambda_{1} & =2.46(1+3.8\mu^*),\\
\Lambda_2&=1.82(1+6.3\mu^*)({\omega}_2/\omega_{\log}), \\
\mu^{*} & =\frac{\mu}{1+\mu\ln(\varepsilon_F/\hbar\omega_D)}.
\end{aligned}}
\label{McMillan}
\end{equation}

In Eq.~{\ref{McMillan}}, $\mu^{*}$ is the Morel-Anderson pseudo-potential describing the screened electron-electron interaction, with
\begin{equation}
\mu=\frac{N(0)}{\pi k_F}\int_0^{2k_F}dqV_C(q)\left[1-\left(\frac{q}{2k_F}\right)^2\right]^{-1/2}
\end{equation}
being the average electron-electron interaction matrix element, $\hbar\omega_D=35 \ \text{meV}$ is the Debye energy of aluminum. And
\begin{equation}
\begin{aligned}
\omega_{\log}&=\exp\left[\langle\ln(\omega)\rangle\right],\\
{\omega}_2&=\sqrt{\langle\omega^2\rangle},
\end{aligned}
\end{equation}
are the average energy with respect to the weight function $\alpha^2F(\omega)\omega$. In principle, it also requires $\lambda\leq10$ and $\mu^*\leq0.2$ to make the McMillan equation valid.

The imaginary axis Eliashberg equation is:
\begin{widetext}
\begin{align}
Z(i\omega_{m})&=1+\frac{\pi k_BT}{\omega_{m}}\sum_{m^{\prime}=-\infty}^{+\infty}\lambda(i\omega_{m}-i\omega_{m^{\prime}})\frac{\omega_{m^{\prime}}}{\sqrt{\omega_{m^{\prime}}^{2}+\Delta^{2}(i\omega_{m^{\prime}})}},\\
Z(i\omega_{m})\Delta(i\omega_{m})&=\pi k_BT\sum_{m^{\prime}=-\infty}^{+\infty}\left[\lambda(i\omega_{m}-i\omega_{m^{\prime}})-u^*\theta(\omega_D-|\omega_{m^{\prime}}|)\right]\frac{\Delta(i\omega_{m^{\prime}})}{\sqrt{\omega_{m^{\prime}}^{2}+\Delta^{2}(i\omega_{m^{\prime}})}}.
\end{align}
\end{widetext}
In the imaginary axis Eliashberg equation, $i\omega_m=(2m-1)k_BT_c$ with $m\in\mathbb{Z}$ is the Matsubara frequency, $\omega_D=35 \  \text{meV}$ is the Debye frequency of aluminum, $\lambda(i\omega_m)$ is the electron-boson interacting constant. $\lambda(i\omega_m)=\lambda_{ph}(i\omega_m)+\lambda_{ex}(i\omega_m)$, where $\lambda_{ph}(i\omega_m)=\lambda^*\frac{\omega_D^2}{\omega^2_D+\omega_m^2}$ is the phonon contribution with $\lambda^*=0.48$ for the aluminum metal\cite{PhysRevB.54.16487}, while $\lambda_{ex}(i\omega_m)=\int_0^\infty d\omega \frac{2\alpha^2 F(\omega)}{\omega^2+\omega_m^2}$ is the exciton contribution where the spectral function can be found in Eq.~\ref{a2f}. $Z$ and $\Delta$ are the renormalization factor and the superconducting gap function, respectively. The imaginary axis Eliashberg equation is solved for temperature $T\in[0.5\ \text{K},4\ \text{K}]$, and the gap function with the first Matsubara frequency $\Delta(i\omega_1)$ is plotted in Figure~\ref{fig3}(b).


\bibliography{Reference}

\end{document}